\documentclass[]{iopart}
\usepackage{graphicx}

\begin{document}

\title[The effect of material parameter ]{\LARGE \bf The effect of material parameter values on the relation between the energy gap width and the scatterer symmetry in two-dimensional phononic crystals}

\author{\bf Izabela \'Sliwa\dag, Maciej Krawczyk\ddag}
\address{\dag\ Nonlinear Optics Division, Adam Mickiewicz University, Umultowska 85, 61-614 Pozna\'n, Poland}
\address{\ddag\ Surface Physics Division, Adam Mickiewicz University, Umultowska 85, 61-614 Pozna\'n, Poland}

\ead{krawczyk@amu.edu.pl}

\date{\today}

\begin{abstract}
The phononic band structures of two-dimensional solid phononic crystals with different lattice and scatterer symmetry are studied numerically,
with three types of lattice (square, triangular and rectangular) and four different scatterer shapes (circle, hexagon, square and rectangle)
considered. $XY$ and $Z$ vibration modes are investigated separately. Two types of phononic crystal are considered: one composed of high-density
rods embedded in a low-density matrix, the other of low-density rods in a high-density matrix. In the former case, lattice type and polarization
being fixed, the broadest gaps are obtained when the symmetry of the rods corresponds to that of the lattice (the shape of a rod is identical
with that of the first Brillouin zone); the largest gap width values are observed in triangular lattice-based crystals (compared to those based
on square and rectangular lattices), the shape of the corresponding first Brillouin zone being closest to a circle. These rules do not apply to
structures in which the density of the rod material is lower than that of the matrix. In this case, when the symmetry of the rods corresponds to
that of the lattice, gaps either fail to appear at all, or are much narrower than in other configurations. The effect of other material
parameter values (such as the longitudinal and transversal velocity values) on the considered relation proves much lesser.
\end{abstract}

\submitto{\JPD}

\maketitle

\section{Introduction}

Elastic wave propagation in periodic structures referred to as
{\em phononic crystals} (PC) has been the subject of intensive
research over the last few years \cite{f1, f2, f3, f4,  f5, f6,
f7}. Composed of elastic centers ({\em e. g.} cylinders, in the two-dimensional case) embedded in a matrix of a different elastic
material and disposed in nodes of a crystal lattice, phononic
crystals show in their band structure energy gaps forbidden to
elastic wave propagation. Due to this property, these structures
have extensive practical applications, especially as so-called
acoustic screens, suppressing waves from a given frequency range
\cite{ekran}. For this reason, it is important to know how to
design structures with the phononic gap as wide as possible.

Beside the physical properties of the PC crystal component
materials, the phononic gap width is found to depend strongly on
the lattice symmetry, as well as on the scatterer shape
\cite{symetria, kwadrat}. This dependence is a common feature of
phononic and photonic crystals \cite{shapes}. In both crystals,
the gap width depends also on the rod orientation with respect to
the lattice, and thus on the relative orientation of the rods as
well \cite{kwadrat, orientacja}.

The effect of scatterer shape and symmetry on the phononic band gap in two-dimensional solid phononic crystals has recently been studied by W.Kuang {\em et al.} \cite{nasza}; the investigated structures consisted of steel rods embedded in epoxy, and thus represented the case of high-density rods in a low-density matrix only. Three types of lattice (triangular, hexagonal and square) and four different scatterer shapes (hexagon, circle, square and triangle) were considered. For a given lattice symmetry, the absolute phononic band gap width was found to be the largest when the shape and orientation of the scatterers corresponded to those of the lattice. Besides, for a given scatterer shape, the gap was found to be the broadest when the lattice had the largest coordination number, as the crystal symmetries in this case were not reduced by the scatterers. Also, the band gap was shown to be controllable by adjusting the scatterer orientation and size: the normalized gap width was found to increase as a result of scatterer rotation if the latter caused the maximum allowed filling fraction to increase; otherwise the normalized gap width would be reduced.
The aim of this study is to show the effect of rod and matrix material parameters (the mass density, in particular) on the relation between the energy gap width and the symmetry of the scatterers and the crystal lattice in two-dimensional phononic crystals. This objective will be realized by investigating the dependence of the gap width on the rod symmetry, with the rods embedded in different sorts of lattice; the two cases: high-density rods in low-density matrix, and low-density rods in high-density matrix, will be considered separately. Four composite types will be examined first: steel/epoxy, C/epoxy, Pb/epoxy and duralumin/epoxy, the epoxy component being always the matrix. Inverse structures, {\em i.e.} epoxy/steel, epoxy/C, epoxy/Pb and epoxy/duralumin, representing the case of low-density rods in a high-density matrix, will be investigated in the next step.
Phononic band gaps will be studied in two-dimensional solid phononic crystals differing in lattice symmetry and rod shape. Specifically, three types of lattice: square ({\em sq}), triangular ({\em tr}) and rectangular ({\em rc}), and four different scatterer shapes: circle (C), hexagon (HX), square (SQ) and rectangle (RC), will be considered. The rod orientation with respect to the lattice is fixed, the symmetry axes of the rods corresponding to those of the crystal lattice with the same symmetry, the configuration allowing to generate the broadest energy gaps in phononic crystals composed of two solids \cite{nasza}. In our investigation of elastic wave propagation in these structures, $XY$ modes (polarized in the wave propagation plane) and $Z$ modes (polarized in the direction perpendicular to the wave propagation) will be examined separately. The effect of rod and matrix material parameters, especially the density ratio between the rod material and the matrix material, on the relation between the energy gap width and the lattice and rod symmetry proves to be substantial for both $Z$ and $XY$ modes.

The paper is divided into four sections. In Section 2, we describe
the plane wave (PW) method used in the band structure calculation;
the numerical results, as well as a table of material parameters,
are presented in Section 3; the results are discussed in Section
4.

\section{Method of calculation}

The phononic band structure is calculated through the elastic
medium wave equation, solved with the use of the plane wave method
\cite{kwadrat, Vasseur}. The plane wave method consists in
transforming a differential wave equation, whose coefficients are
periodic functions of two-dimensional position vector
$\vec{r}=(x_{1},x_{2})$, into the reciprocal space. For a
nonhomogeneous elastic medium the wave equation reads
\cite{Landau}:
\begin{eqnarray}\label{eq1}
\rho (\vec{r}) \frac{\partial^{2} u_{i}}{\partial t^{2}}=\nabla \cdot \left[\rho (\vec{r}) c_{t}^{2}(\vec{r})\nabla u_{i}\right]+\nabla \cdot \left[\rho (\vec{r})c_{t}^{2}(\vec{r}) \frac{\partial \vec{u}}{\partial x_{i}}\right]\\
+\frac{\partial}{\partial x_{i}}\left[\left(\rho (\vec{r})c_{l}^{2}(\vec{r})-2\rho (\vec{r})c_{t}^{2}(\vec{r})\right)\nabla \cdot \vec{u}\right],\nonumber
\end{eqnarray}
where $c_{t}$ and $c_{l}$ are the elastic wave transversal and
longitudinal velocities, respectively; $\rho $ denotes mass
density, and $\vec{u}(\vec{r},t)$ stands for the time and
space-dependent deflection vector, describing the medium molecules
deflection from the balance point, and defined by its three
components $u_{i}$ ($i=1, 2, 3$) in the Cartesian coordinate
system $(0x_{1}x_{2}x_{3})$.

The structure is assumed to be homogeneous along the $x_{3}$ axis,
which reduces the problem to two dimensions. If the wave
propagation is confined to the phononic crystal periodicity plane
(the vector $\vec{r_{\parallel }}=(x_{1}, x_{2})$ plane),
(\ref{eq1}) splits into two independent equations \cite{Vasseur}:
\begin{itemize}
\item an equation describing $XY$ modes, or oscillations polarized in plane $(x_{1}0x_{2})$:
\begin{eqnarray}\label{eq3}
\rho (\vec{r_{\parallel }})\frac{\partial^{2} u_{i}}{\partial t^{2}}=\nabla_{T} \cdot \left[\rho (\vec{r_{\parallel }})c_{t}^{2}(\vec{r_{\parallel }})\nabla_{T}u_{i}\right]+\nabla_{T} \cdot
\left[\rho (\vec{r_{\parallel }})c_{t}^{2}(\vec{r_{\parallel }})\frac{\partial \vec{u_{T}}}{\partial x_{i}} \right]\nonumber \\
+\frac{\partial}{\partial x_{i}}\left[\left(\rho (\vec{r_{\parallel }})c_{l}^{2}(\vec{r_{\parallel }})-2\rho (\vec{r_{\parallel }})c_{t}^{2}(\vec{r_{\parallel }})\right)\nabla_{T} \cdot \vec{u_{T}}\right],
\end{eqnarray}
where
\begin{eqnarray*}
\vec{u_{T}}=u_{1}\vec{e_{1}}+u_{2}\vec{e_{2}},\nonumber \\
\nabla_{T}=\frac{\partial}{\partial x_{1}}\vec{e_{1}}+\frac{\partial}{\partial x_{2}}\vec{e_{2}}\nonumber
\end{eqnarray*}
for $i=1, 2$. This equation can be re-written with Lame
coefficients:
\begin{equation}\label{equation3}
\frac{\partial^{2} u_{i}}{\partial t^{2}}=\frac{1}{\rho }\left\{\frac{\partial}{\partial x_{i}}\left(\lambda \frac{\partial u^{l}}{\partial
x_{l}}\right) +\frac{\partial}{\partial x_{l}}\left[\mu \left(\frac{\partial u_{i}}{\partial x_{l}}+\frac{\partial u_{l}}{\partial
x_{i}}\right)\right]\right\},
\end{equation}
where
\begin{displaymath}
\begin{array}{ccl}
\mu & = & \rho c_{t}^{2},\\
\lambda & = & \rho (c_{l}^{2}-2c_{t}^{2})\;\;\;,
\end{array}
\end{displaymath}
and \item an equation describing $Z$ modes, or oscillations
polarized along the rod axis, {\em i. e.} along vector
$\vec{e_{3}}$ ($\vec{e_{3}}\perp \vec{k}$):
\begin{equation}\label{eq4}
\rho (\vec{r_{\parallel }})\frac{\partial^{2} u_{3}}{\partial t^{2}}=\nabla \cdot \left[\rho (\vec{r_{\parallel}})c_{t}^{2}(\vec{r_{\parallel }})\nabla u_{3}\right].
\end{equation}
\end{itemize}

\begin{figure}[h]
\begin{center}
\includegraphics[height=11cm ]{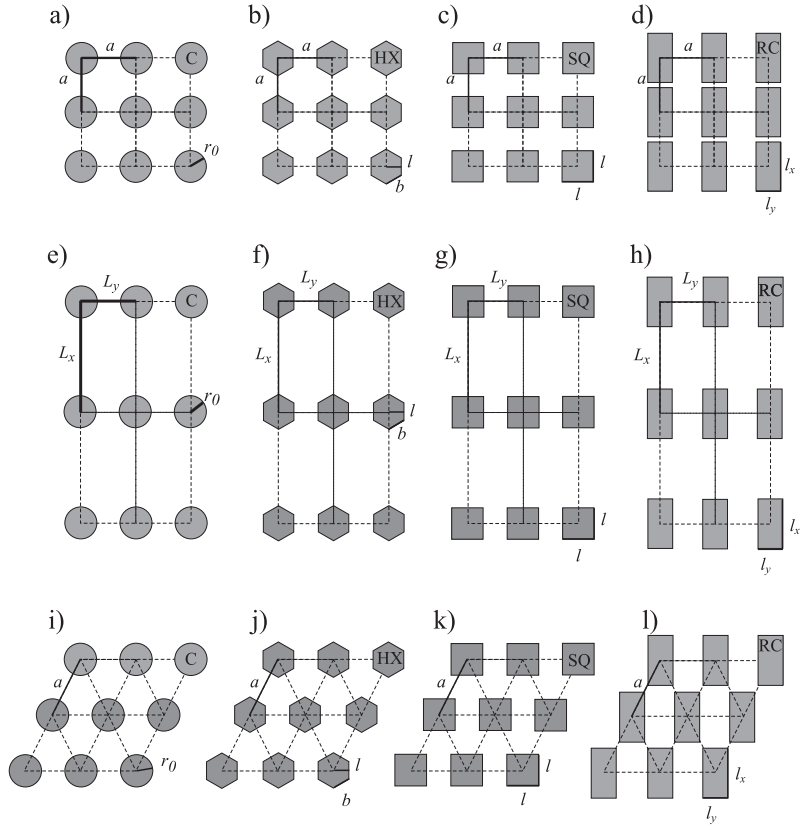}
\caption{\footnotesize {\sf . Models of two-dimensional phononic crystals based on different types of lattice and with different rod
cross-section shapes: square ({\em sq}) lattice-based crystals (the lattice constant being $a$) with rod cross section in the shape of: circle C
with radius $r_{0}$ (a), regular hexagon HX with side length $b$ (b), square SQ with side length $l$ (c), and rectangle RC with side lengths
$l_{1}$ and $l_{2}$ (d); rectangular ({\em rc}) lattice-based crystals (the lattice constants being $L_{1}$ and $L_{2}$ with rod cross section
in the shape of: circle (e), hexagon (f), square (g) and rectangle (h); triangular ({\em tr}) lattice-based crystals (the lattice constant being
a) with rod cross section in the shape of: circle (i), hexagon (j), square (k) and rectangle (l).}}\label{Rys.1}
\end{center}
\end{figure}

As the coefficients in equations of motion (\ref{equation3}) and (\ref{eq4}) are periodic functions, through Bloch's theorem, the deflection vector, $\vec{u}(\vec{r},t)$, can be expressed as \cite{kwadrat}:
\begin{equation}\label{eq5}
\vec{u}(\vec{r_{\parallel }},t)=e^{i(\vec{k}\cdot \vec{r_{\parallel }}-\omega
t)}\sum_{\vec{G}} \vec{u}_{\vec{k}}(\vec{G})e^{i\vec{G}\cdot
\vec{r_{\parallel }}},
\end{equation}
$\vec{k}$ being a two-dimensional Bloch vector, and $\vec{G}$ denoting a reciprocal lattice vector. The reciprocal lattice vectors, $\vec{G}$, for the lattices considered in this paper (Figure \ref{Rys.1}) are defined as follows:
\begin{itemize}
\item for a square lattice
\begin{equation}\label{eq5'}
\vec{G}=G_{1}\vec{e_{1}}+G_{2}\vec{e_{2}}=\frac{2\pi }{a}\left(n_{1}\vec{e_{1}}+n_{2}\vec{e_{2}}\right),
\end{equation}
\item for a triangular lattice
\begin{equation}
\vec{G}=G_{1}\vec{e_{1}}+G_{2}\vec{e_{2}}=\frac{2\pi }{a\sqrt{3}}\left[n_{1}\sqrt{3}\vec{e_{1}}+\left(-n_{1}+2n_{2}\right)\vec{e_{2}}\right],
\end{equation}
\item for a rectangular lattice
\begin{equation}
\vec{G}=G_{1}\vec{e_{1}}+G_{2}\vec{e_{2}}=\frac{2\pi }{L_{1}L_{2}}\left[n_{1}L_{2}\vec{e_{1}}+n_{2}L_{1}\vec{e_{2}}\right],
\end{equation}
\end{itemize}
$n_{1}$ and $n_{2}$ being integers, $a$ being a lattice constant in the case of square and triangular lattices, and $L_{1}$ and $L_{2}$ being the rectangular lattice constants along the $x_{1}$ and $x_{2}$ directions, respectively.

Those of the material parameters which are periodic functions can
be Fourier-expanded. For $Z$ modes, described by (\ref{eq4}),
Fourier expansion can be applied to the inverse of mass density,
$\rho^{-1}$, and to coefficient $\mu $; in (\ref{equation3}),
describing $XY$ modes, the Fourier-expandable parameters are
$\rho^{-1}$, $\mu $ and $\lambda $:
\begin{equation}\label{eq6}
\rho^{-1} (\vec{r_{\parallel }})=\sum_{\vec{G}} \rho^{-1} (\vec{G})e^{i\vec{G}\cdot \vec{r_{\parallel }}},
\end{equation}
\begin{equation}\label{eq61}
\mu (\vec{r_{\parallel }})=\sum_{\vec{G}} \mu (\vec{G})e^{i\vec{G}\cdot \vec{r_{\parallel }}},
\end{equation}
\begin{equation}\label{eq62}
\lambda (\vec{r_{\parallel }})=\sum_{\vec{G}} \lambda (\vec{G})e^{i\vec{G}\cdot \vec{r_{\parallel }}}.
\end{equation}

Coefficients $\rho^{-1} (\vec{G})$, $\mu (\vec{G})$ and $\lambda
(\vec{G})$ in the above expansions are calculated from the inverse
Fourier transformation. All these coefficients will have the same
form; for example, the inverse of mass density will read:
\begin{equation}\label{eq7}
\rho^{-1} (\vec{G})=\frac{1}{S}\int \int \rho^{-1} (\vec{r_{\parallel }})e^{-i\vec{G}\cdot \vec{r_{\parallel }}}d\vec{r_{\parallel }},
\end{equation}
the integration covering the two-dimensional unit cell surface,
$S$. When $\vec{G}=0$, (\ref{eq7}) defines the mean value:
\begin{equation}\label{eq8}
\rho^{-1} (\vec{G} =0)=\bar{\rho^{-1} } \equiv \rho_{A}^{-1}f+\rho_{B}^{-1}(1-f),
\end{equation}
where $\rho_{A}$ is the mass density value in the rod, $\rho_{B}$
is its matrix counterpart, and $f=S_{r}/S$ denotes the filling
fraction ($S_{r}$ being the rod cross-section area). When
$\vec{G}\neq 0$, (\ref{eq7}) becomes:
\begin{equation}\label{eq9}
\rho^{-1} (\vec{G}\neq
0)=\left(\rho_{A}^{-1}-\rho_{B}^{-1}\right)I(\vec{G})=\Delta
\rho^{-1} I(\vec{G}),
\end{equation}
where $\Delta \rho^{-1} =\rho_{A}^{-1}-\rho_{B}^{-1}$, and $I(\vec{G})$ is a structural factor determined by the rod shape:
\begin{equation}\label{eq10}
I(\vec{G})\equiv \frac{1}{S} \int \int_{S_{r}} e^{-i\vec{G}\cdot
\vec{r_{\parallel }}}d\vec{r_{\parallel }}.
\end{equation}
The integration in (\ref{eq10}) is performed over the rod cross
section. For the four types of rods (Figure \ref{Rys.1}), with
circular, hexagonal, rectangular or square cross section, the
integration results in the following formulae ($I_{C}(\vec{G})$,
$I_{HX}(\vec{G})$, $I_{RC}(\vec{G})$ and $I_{SQ}(\vec{G})$
denoting the respective structural factors):
\begin{itemize}
\item for cylindrical rods:
\begin{equation}\label{eq13}
I_{C}(\vec{G})=2f\frac{J_{1}(Gr_{0})}{Gr_{0}},
\end{equation}
where $J_{1}$ is a first-order Bessel function, $r_{0}$ is the cylinder cross-section radius, and filling fraction $f$ is defined as follows:
\begin{displaymath}
f=\frac{\pi r_{0}^{2}}{a^{2}},\;\;\; 0\leq f\leq \frac{\pi }{4}\;\;\;\mbox{for a square lattice},
\end{displaymath}
\begin{displaymath}
f=\frac{4\pi r_{0}^{2}}{a^{2}\sqrt{3}}, \;\;\;0\leq f\leq \frac{\pi }{\sqrt{3}} \;\;\; \mbox{for a triangular lattice},
\end{displaymath}
and
\begin{displaymath}
f=\frac{\pi r_{0}^{2}}{L_{1}L_{2}} \;\;\;\;\;\;\;\;\mbox{for a rectangular lattice};
\end{displaymath}
\item for hexagonal rods:
\begin{equation}\label{eq15}
\fl I_{HX}(\vec{G})=
\left\{
\begin{array}{l}
\frac{2f}{3G_{1}l}\left[\sin (G_{1}l)+\frac{1-\cos (G_{1}l)}{G_{1}l}\right],\;\; G_{1}\neq 0, \; G_{2}=0, \\
\frac{f}{3G_{1}l}\left[\frac{1-\cos (2G_{1}l)}{2G_{1}l}+\sin (2G_{1}l)\right],\;\;\;|G_{1}|=|G_{2}|/\sqrt{3},\;\;\; G_{2}\neq 0, \\
\frac{f}{G_{2}l}\left[\frac{\cos (G_{2}l/\sqrt{3}-G_{1}l)-\cos (2G_{2}l/\sqrt{3})}{\left(G_{2}+G_{1}\sqrt{3}\right)l}+\frac{\cos (G_{2}l/\sqrt{3}+G_{1}l)-\cos (2G_{2}l/\sqrt{3})}{\left(G_{2}-G_{1}\sqrt{3}\right)l}\right],\\
\begin{array}{rrr}
  &  & |G_{1}|\neq |G_{2}|/\sqrt{3}, \;\;\; |G_{2}|\neq 0,
\end{array}
\end{array}
\right.
\end{equation}
where $l=b\sqrt{3}/2$, $b$ denoting the regular hexagon side length; the corresponding filling fraction is:
\begin{displaymath}
f=\frac{3b^{2}\sqrt{3}}{2a^{2}},\;\;\;0\leq f\leq \frac{3\sqrt{3}}{8}\;\;\; \mbox{for a square lattice},
\end{displaymath}
\begin{displaymath}
f=\frac{3b^{2}}{a^{2}},\;\;\; 0\leq f\leq 1\;\;\; \mbox{for a triangular lattice},
\end{displaymath}
and
\begin{displaymath}
f=\frac{3b^{2}\sqrt{3}}{2L_{1}L_{2}}\;\;\;\;\;\;\;\;\;\mbox{for a rectangular lattice};
\end{displaymath}
\item for rectangular rods:
\begin{equation}\label{eq15'}
I_{RC}(\vec{G})=
\left\{
\begin{array}{l}
\frac{2l_{1}}{G_{2}}\sin (\frac{l_{2}G_{2}}{2}),\;\; G_{1}=0, \; G_{2}\neq 0, \\
\frac{2l_{2}}{G_{1}}\sin (\frac{l_{1}G_{1}}{2}),\;\;\; G_{2}=0, \; G_{1}\neq 0, \\
\frac{4}{G_{1}G_{2}}\sin (\frac{l_{1}G_{1}}{2})\sin (\frac{l_{2}G_{2}}{2}),\;\; G_{1}\neq 0,\; G_{2}\neq 0,\\
\end{array}
\right.
\end{equation}
where $l_{1}$ and $l_{2}$ are the rectangle side lengths, and filling fraction $f$ reads:
\begin{displaymath}
f=\frac{l_{1}l_{2}}{a^{2}}\;\;\; \;\;\;\;\;\mbox{for a square lattice},
\end{displaymath}
\begin{displaymath}
f=\frac{2l_{1}l_{2}}{a^{2}\sqrt{3}}\;\;\;\;\;\;\;\; \mbox{for a triangular lattice},
\end{displaymath}
and
\begin{displaymath}
f=\frac{l_{1}l_{2}}{L_{1}L_{2}}\;\;\;\;\;\;\;\;\mbox{for a rectangular lattice};
\end{displaymath}
\item the structural factor for square rods, $I_{SQ}(\vec{G})$, is deduced from  (\ref{eq15'}) through putting $l_{1}=l_{2}=l$:
\begin{equation}\label{eq15bis}
I_{SQ}(\vec{G})=I_{RC}(\vec{G}) \;\;\; \mbox{with} \;\;\; l_{1}=l_{2}=l.
\end{equation}
\end{itemize}
The maximum filling fraction values for rectangular lattice and rods depend on the rectangle side length ratio, $l_{x}/l_{y}$ and the lattice constant ratio, $L_{x}/L_{y}$.

The substitution of (\ref{eq5}), (\ref{eq6} - \ref{eq62}), (\ref{eq8}) and (\ref{eq9}) into (\ref{equation3}) leads to an infinite system of algebraic equations for $XY$ mode eigenvalues $\omega (\vec{k})$:
\begin{equation}\label{eq11}
\fl \begin{array}{ccl}
\omega^{2}u_{\vec{k}+\vec{G}}^{i} & = & \sum_{\vec{G'}}\left\{\sum_{l, \vec{G''}}\rho^{-1}(\vec{G}-\vec{G''})\left[\lambda (\vec{G''}-\vec{G'})(\vec{k}+\vec{G''})_{l}(\vec{k}+\vec{G''})_{i} \nonumber \right. \right. \\
  &  & \left. +\mu (\vec{G''}-\vec{G'})(\vec{k}+\vec{G'})_{i}(\vec{k}+\vec{G''})_{l}\right]u_{\vec{k}+\vec{G'}}^{l} \nonumber\\
  &  & \left. +\sum_{\vec{G''}}\left[\rho^{-1}(\vec{G}-\vec{G''})\mu (\vec{G''}-\vec{G'})\sum_{j}(\vec{k}+\vec{G'})_{j}(\vec{k}+\vec{G''})_{j}\right]u_{\vec{k}+\vec{G'}}^{i}\right\},
\end{array}
\end{equation}
where $i,j,l = 1,2$.
Analogically, the following $Z$-mode eigenproblem is deduced from (\ref{eq4}):
\begin{eqnarray}\label{eq12}
\fl \omega^{2}u_{\vec{k}}(\vec{G'})=\sum_{\vec{G'}, \vec{G''}}\rho^{-1} (\vec{G}-\vec{G''})\mu(\vec{G''}-\vec{G'})(\vec{k}+\vec{G'})\cdot
(\vec{k}+\vec{G''})u_{\vec{k}}(\vec{G'}).
\end{eqnarray}

\section{Numerical results and discussion}

The material parameter values  ($\rho $, $c_{l}$, $c_{t}$) in all the materials involved (C, epoxy, steel, Pb and duralumin), as well as the rod/matrix parameter value contrasts ($\Delta \rho =|\rho_{A}-\rho_{B}|$, $\Delta c_{l}=|c_{l}^{A}-c_{l}^{B}|$, $\Delta c_{t}=|c_{t}^{A}-c_{t}^{B}|$), are specified in Table \ref{tab.1}.
\begin{table}[thp]
\caption{\footnotesize {\sf Mass density, transversal velocity
and longitudinal velocity values in the component materials of the
studied phononic crystals. Parameter value contrasts between the
rods and the matrix ($\Delta \rho $, $\Delta c_{l}$ and $\Delta
c_{t}$) are specified as well.}}\label{tab.1}
\begin{indented}
\item[]\begin{tabular}{ccccccc} \hline \hline \br
{\em Material} & $\rho $ & $\Delta \rho $ & $c_{t}$ & $\Delta c_{t}$ & $c_{l}$ & $\Delta c_{l}$\\
  & $[kg/m^{3}]$ & $[kg/m^{3}]$ & $[m/s]$ & $[m/s]$ & $[m/s]$ & $[m/s]$\\ \hline
C & $1750$ &  & $7100$ &  & $13310$ \\
epoxy & $1200$ & $550$ & $1160$ & $5940$ & $2830$ & $10480$ \\ \hline
steel & $7800$ &  & $3220$ &  & $5940$ \\
epoxy & $1200$ & $6600$ & $1160$ & $2060$  & $2830$ & $3110$ \\ \hline
Pb & $11357$ &  & $860$ &  & $2158$ \\
epoxy & $1200$ & $10157$ & $1160$ & $300$ & $2830$ & $672$  \\ \hline
duralumin & $2799$ &  & $3095$ &  & $6342$ \\
epoxy & $1200$ & $1599$ & $1160$ & $1935$  & $2830$ & $3512$ \\ \hline 
\br
\end{tabular}
\end{indented}
\end{table}

The material parameter value contrasts vary substantially with
structure composition, $\Delta \rho $ ranging from 550 to 10157,
$\Delta c_{t}$ from 300 to 5940, and $\Delta c_{l}$ from 672 to
10480. This variety increases the generality of our conclusions, drawn
from the results obtained for different phononic crystals, and
regarding the effect of the rod and lattice symmetry on the energy gap
width in their elastic wave spectrum.

In the numerical solution of (\ref{eq11}) and (\ref{eq12}),
$I(\vec{G})$ is substituted by $I_{C}(\vec{G})$ for cylindrical
rods (\ref{eq13}), $I_{HX}(\vec{G})$ for hexagonal rods
(\ref{eq15}),  $I_{RC}(\vec{G})$ for rectangular rods
(\ref{eq15'}) and $I_{SQ}(\vec{G})$ for square ones
(\ref{eq15bis}). The number of reciprocal lattice vectors involved
in Fourier expansions (\ref{eq6})-(\ref{eq62}) is limited by
condition $-N\leq n_{1}$, $n_{2}\leq +N$, confining the reciprocal
lattice vector component values $n_{1}$ and $n_{2}$ to interval
$[-N, +N]$. From the resulting finite system of algebraic
equations for $2N+1$ coefficients $u_{i}(G)$, $2N+1$ allowed
frequencies $\omega (\vec{k})$ are deduced for each wave vector
from the first Brillouin zone. All the results presented in this
paper were obtained for $N=7$. Some dispersion relations were
calculated for $N=10$ as well, and found to be in very good
conformity with those obtained for $N=7$, regardless of the rod
shape.

Among the rod shapes considered, only the rectangular cross-section leaves some arbitrariness, which lies in the rectangle side ratio, $l_{1}/l_{2}$. In order to find the optimum side ratio, {\em i.e.} that corresponding to the broadest energy gap, in a rectangular lattice-based crystal with rectangular rods, the dependence of the gap width on the filling fraction was investigated for the following values of $l_{1}/l_{2}$: $1/2$, $1$, $2$ and $3$, the lattice constant ratio being fixed at $L_{1}/L_{2}= 2$. Some of the results obtained for the steel/epoxy crystal are shown in Table \ref{tab.4}.
Energy gap width values, as well as frequency values, are specified, throughout this study, in the units of reduced frequency,$\Omega $, defined as follows:
\begin{equation}\label{eq101}
\Omega =\frac{\omega a_{0}}{4\pi \sqrt{\bar{c_{t}}}},
\end{equation}
$a_{0}$ being the distance between two neighboring lattice nodes, and $\bar{ c_{t}}$ denoting the mean between the transversal velocities in materials A and B. The energy gap is found to be the broadest for the side ratio equal to the constant lattice ratio, $l_{1}/l_{2}=2$ in the case considered. Therefore, we shall henceforth confine our analysis to rectangular rods with side ratio $l_{1}/l_{2}=2$ and a rectangular lattice with constant lattice ratio $L_{1}/L_{2}=2$. The specific value of the constant lattice ratio, as well as that of the side length ratio, is of no importance for this study as long as both ratios are equal, which means that the symmetry of the rods corresponds to that of the lattice.
\begin{table}
\caption{\footnotesize {\sf The width values of the broadest energy gap in steel/epoxy crystals consisting of rectangular rods of different
$l_{1}/l_{2}$ values disposed in nodes of a rectangular lattice ($L_{1}/L_{2}=2$).}}\label{tab.4}
\begin{indented}
\item[]\begin{tabular}{ccc} \br
$l_{1}/l_{2}$ & \em $Z$ mode & \em $XY$ mode \\ 
\mr
$1/2$ & $0$ & $0$\\ 
$1$ & $0$ & $0$\\ 
$2$ & $0.66$ & $0.50$\\ 
$3$ & $0.47$ & $0.22$\\
\br
\end{tabular}
\end{indented}
\end{table}
Once the component materials, the rod shape and the crystallographic lattice type are fixed, in the next step of our investigation of the effect of the symmetry relation between the rod and the lattice on the width of a phononic energy gap, we shall specify the filling fraction value corresponding to the maximum gap width. For each material composition, the reduced frequency is plotted against the filling fraction, $f$, on the basis of (\ref{eq11}) and (\ref{eq12}), with three possible lattice types (square, triangular and rectangular), four rod cross-section shapes (square, circle, regular hexagon and rectangle) and two polarizations considered; the aim is to find, in each case, the optimum filling fraction value, corresponding to the maximum width of the broadest gap.

\begin{figure}[h]
\begin{center}
\includegraphics[height=5cm ]{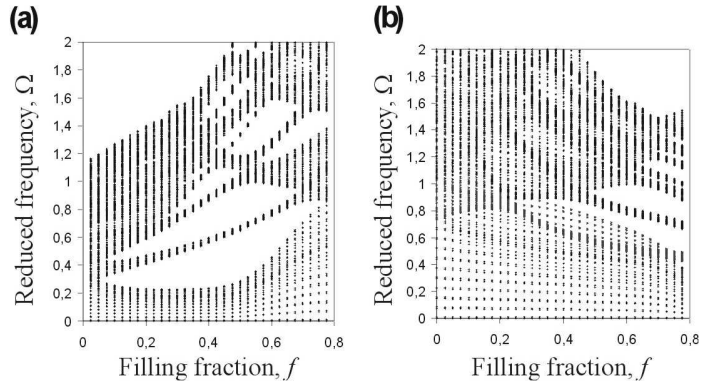}
\caption{\footnotesize {\sf $Z$-mode reduced frequency plotted against the filling fraction, $f$, for a steel/epoxy crystal with cylindrical
rods disposed in square lattice nodes (a), and its epoxy/steel counterpart (b).}}\label{Rys.3}
\end{center}
\end{figure}

Figure \ref{Rys.3} shows some examples of phononic crystal energy bands plotted against the filling fraction in two cases: (a) in a steel/epoxy crystal, representing the case of high-density rods in a low-density matrix ($\rho_{B}<\rho_{A}$), and (b) in an epoxy/steel structure, with low-density rods in a high-density matrix ($\rho_{B}>\rho_{A}$). These two qualitatively different cases bring out the significant effect of the material parameter values on the relations obtained. In the case of high-density rods in a low-density matrix, the optimum filling fraction value for the broadest gap (found between the first and the second band) is f = 0.48, the gap center shifting up the frequency scale as $f$ increases    (Figure \ref{Rys.3}(a)). In the other case, the gaps are narrower, their centers shift down as $f$ increases, and the optimum filling fraction value is much higher that in the case considered previously (Figure \ref{Rys.3}(b)). Therefore, in the following sections of this paper, the cases of high-density rods in a low-density matrix and low-density rods in a high-density matrix will be considered separately.

\subsection{The case of high-density rods in a low-density matrix}

In this paragraph we shall focus on structures in which the matrix material is assumed to be epoxy; due to the lowest mass density of this material, inequality $\rho_{A}>\rho_{B}$, conducive to the appearance of wider energy gaps \cite{Vasseur}, is fulfilled in each case. The width values of the broadest energy gaps obtained in each of the examined structures, and corresponding to the optimum filling fraction values, are specified in Table \ref{tab.3}.
\begin{table}
\caption{\footnotesize {\sf The width values of the broadest energy gap in the examined phononic crystals, in reduced frequency units. {\em sq} denotes square lattice, {\em tr} triangular lattice, {\em rc} rectangular lattice, {\em C} cylindrical rods, {\em HX}
hexagonal rods, {\em SQ} square rods, and {\em RC} rectangular rods.}}\label{tab.3}
\begin{indented}
\item[]\begin{tabular}{lcccccc}
\br
\em Material & \em Polarization & \em Lattice & \em C & \em HX & \em SQ & \em RC\\ 
\mr
  &  & \em sq & $0.37$ & $0.36$ & $0.64$ & $0.20$\\
  & $Z$ & \em tr & $0.77$ & $0.80$ & $0.75$ & $0.31$\\ 
steel  &  & \em rc & $0.04$ & $0.05$ & $0$ & $0.47$\\ 
\cline{2-7}
/epoxy  &  & \em sq & $0.57$ & $0.51$ & $0.74$ & $0.06$\\ 
  & $XY$ & \em tr & $0.86$ & $0.86$ & $0.75$ & $0.11$\\ 
  &  & \em rc & $0.03$ & $0$ & $0$ & $0.35$\\ \hline
  &  & \em sq & $0.21$ & $0.16$ & $0.30$ & $0.02$ \\
  & $Z$ & \em tr & $0.49$ & $0.55$ & $0.48$ & $0.11$\\
C  &  & \em rc & $0$ & $0.16$ & $0$ & $0.27$\\
\cline{2-7}
/epoxy &  & \em sq & $0.13$ & $0.11$ & $0.37$ & $0$\\
  & $XY$ & \em tr & $0.47$ & $0.49$ & $0.40$ & $0$\\
  &  & \em rc & $0$ & $0$ & $0$ & $0.09$\\ \hline
  &  & \em sq & $0.35$ & $0.35$ & $0.35$ & $0.24$\\
  & $Z$ & \em tr & $0.47$ & $0.47$ & $0.46$ & $0.18$ \\
Pb &  & \em rc & $0.08$ & $0$ & $0.09$ & $0.18$\\
\cline{2-7}
/epoxy &  & \em sq & $0.32$ & $0.32$ & $0.32$ & $0.16$\\
  & $XY$ & \em tr & $0.37$ & $0.37$ & $0.36$ & $0.27$\\
  &  & \em rc & $0$ & $0$ & $0$ & $0.07$\\ \hline
  &  & \em sq & $0.21$ & $0.20$ & $0.35$ & $0.07$\\
  & $Z$ & \em tr & $0.48$ & $0.51$ & $0.44$ & $0.21$\\
duralumin &  & \em rc & $0$ & $0$ & $0$ & $0.18$\\
\cline{2-7}
/epoxy &  & \em sq & $0.29$ & $0.25$ & $0.37$ & $0$\\
  & $XY$ & \em tr & $0.47$ & $0.47$ & $0.40$ & $0.06$\\
  &  & \em rc & $0$ & $0$ & $0$ & $0.08$\\
\br
\end{tabular}
\end{indented}
\end{table}

The results shown in Table \ref{tab.3} indicate that, among the lattice types considered here, it is the triangular lattice ({\em i.e.} the one whose first Brillouin zone has a shape closest to a circle) that allows to obtain energy gaps which are by far the broadest. Besides, the gaps obtained in each of the phononic crystals covered by Table \ref{tab.3} are found to be the broadest in structures in which the symmetry of the rods corresponds exactly to that of the lattice, which confirms the results reported in \cite{nasza}. The results obtained in the case of triangular lattice can be expressed by the following inequality, for both $Z$ and $XY$ modes:
\begin{equation}\label{eq16}
\Delta \Omega_{HX} \geq \Delta \Omega_{C} \geq \Delta \Omega_{SQ} \geq \Delta \Omega_{RC}, \label{omega1}
\end{equation}
where $\Delta \Omega_{SQ}$ is the gap width in the square-rod composite, $\Delta \Omega_{C}$ is the corresponding value in the cylinder-rod structure, $\Delta \Omega_{HX}$ denotes the gap width in the composite with hexagonal rods and $\Delta \Omega_{RC}$ refers to the case of rectangular rods.

The results obtained in the case of square lattice, for both $Z$ and $XY$ modes, can be expressed as follows:
\begin{equation}\label{eq17}
\Delta \Omega_{SQ} \geq \Delta \Omega_{C} \geq \Delta \Omega_{HX}>\Delta \Omega_{RC}. \label{omega2}
\end{equation}
In the case of rectangular lattice, the results obtained cannot be expressed by just one inequality; however, the broadest gaps are found to appear for rectangular rods.

The results obtained for steel/epoxy, C/epoxy and duralumin/epoxy structures are qualitatively the same; in the case of $Z$ modes, all the weak inequalities in (\ref{omega1}) and (\ref{omega2}) can be replaced with strict ones. However, the results obtained for Pb/epoxy are different: as long as the lattice type is fixed, gaps appearing at various rod shapes (cylindrical, hexagonal and square) are found to be equal. The optimum filling fraction value lies within the range from $0.460$ to $0.675$ in each of the first three cases, but is found to be below $f = 0.375$, and thus relatively low, in the case of the Pb/epoxy composition; this means that the matrix material is then the dominant one in the Pb/epoxy crystal.  The described properties of the energy gap width in the Pb/epoxy crystal are due to different contrasts of longitudinal and transversal velocity values in the component materials: the relations between the $c_{l}$ and $c_{t}$ values in the rods and in the matrix are reversed with respect to the other compositions ($c_{tA}<c_{tB}$ and $c_{lA}<c_{lB}$), while the density relation is the same ($\rho_{A}>\rho_{B}$). The gaps are found to be narrowest in the C/epoxy composite, due to the lowest density contrast, $\Delta \rho=550$, between the component materials.

\subsection{The case of low-density rods in a high-density matrix}

In order to check whether the results discussed above depend on the rod and matrix density values, similar computations of the phononic gap width values for different rod symmetries were performed for epoxy/steel, epoxy/C, epoxy/Pb and epoxy/duralumin crystals, representing the case of low-density rods in a high-density matrix. Providing an example, Figure \ref{Rys.3}(b) shows the reduced frequency plotted against the filling fraction on the basis of the results obtained in the epoxy/steel crystal with cylindrical rods embedded in a square lattice.

\begin{figure}[h]
\begin{center}
\includegraphics[height=8cm ]{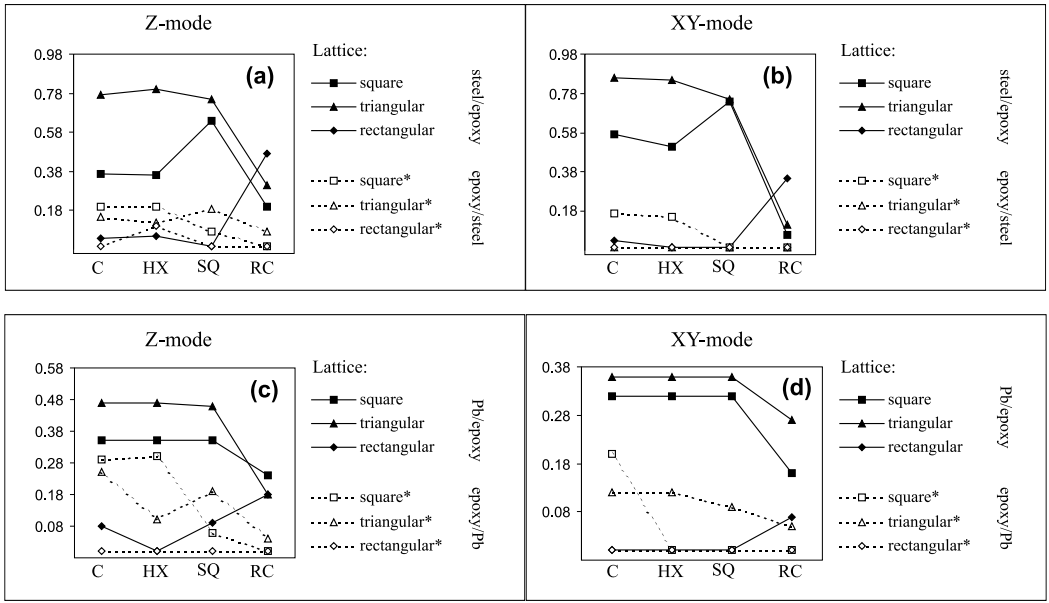}
\caption{\footnotesize {\sf The energy gap width (in the units of reduced frequency $\Omega $) plotted against the rod shape (C, HX, SQ, RC) in
two-dimensional phononic crystals; (a) and (b) show results obtained in steel/epoxy and epoxy/steel crystals; (c) and (d) present those obtained
for Pb/epoxy and epoxy/Pb compositions. $Z$ and $XY$ modes are considered separately, the results being depicted in (a), (c) and (b), (d),
respectively. C denotes cylindrical rods, HX hexagonal rods, SQ square rods, and RC rectangular rods; the rods are disposed in nodes of a
square, a triangular or a rectangular lattice, as indicated by symbols explained in the legend. The asterisk in the legend refers to the
epoxy/steel structures in (a) and (b), and to epoxy/Pb structures in (c) and (d).}}\label{Rys.6}
\end{center}
\end{figure}

In Figure \ref{Rys.6} the gap width is plotted against the rod shape in steel/epoxy and epoxy/steel crystals (Figure \ref{Rys.6} (a) and (b)) as well as in Pb/epoxy and epoxy/Pb crystals (Figure \ref{Rys.6} (c) and (d));  $Z$ and $XY$ modes are considered separately (Figure \ref{Rys.6} (a),(c) and \ref{Rys.6} (b),(d), respectively). The plots obtained for the remaining two compositions, {\em i.e.} for C/epoxy and duralumin/epoxy, are qualitatively the same as those obtained for the steel/epoxy crystal, and thus are not depicted in the figures presented in this paper. The plots shown in Figure \ref{Rys.6} indicate clearly that, regardless of polarization, relations (\ref{omega1})-(\ref{omega2}), obtained for crystals with high-density rods in a low-density matrix (represented by the solid lines), do not apply to the inverse structures (dashed lines), in which the density of the rod material is lower that that of the matrix. In triangular lattice-based crystals, hexagonal rods are found to generate narrower gaps than cylindrical or square rods. In square lattice-based structures, square rods are found to generate no gaps at all (though other rod shapes do generate them); also in rectangular lattice-based structures no gaps are generated by rectangular rods. Therefore, the inversion of the material composition clearly implies a radical change in the relations between the gap width and the scatterer symmetry: in none of the cases considered, for a given lattice type, the broadest gaps are found to appear when the symmetry of the rods corresponds to that of the lattice. Besides, the gaps obtained in such inverse structures are much narrower (four times in the epoxy/steel crystal, and twice in the epoxy/Pb structure) than those observed in the corresponding crystal with high-density rods in a low-density matrix. As similar results were obtained for all the compositions considered (also in the epoxy/Pb crystal, with different longitudinal and transversal velocity contrasts), the strongest impact on the considered relations can be expected to come from the mass density.

\section{Conclusion}

As indicated by the results presented in the preceding section, covering two classes of phononic crystals: those with high-density rods embedded in a low-density matrix, and those with low-density rods in a high-density matrix, material parameter values have much of an impact on the relation between the energy gap width and the lattice and rod symmetry; the effect of the mass density was found to be especially significant. In structures with high-density rods in a low-density matrix, the gaps are found to be the broadest when the symmetry of the rods corresponds to that of the lattice, i.e. for hexagonal rods in triangular lattice-based crystals, square rods in square lattice-based crystals, and rectangular rods in rectangular lattice-based crystals, as already demonstrated in \cite{nasza}. However, this rule will cease to apply when the density of the rod material becomes lower than that of the matrix. In this case, when the symmetry of the rods corresponds to that of the lattice, gaps either fail to appear at all, or are much narrower than in other configurations. Gaps obtained in such inverse structures are always much narrower than those appearing in crystals with high-density rods in a low-density matrix. This is due to the fact that low-density rods embedded in a high-density matrix are weak scattering centers: their wave scattering effect is hardly noticeable and much lesser than that produced by high-density rods in a low-density matrix. As regards the longitudinal and transversal velocity values ($c_{l}$ and $c_{t}$) in the component materials, the effect of these parameters on the investigated relations proves less significant.

\ack The authors are indebted to H. Puszkarski for suggesting this study and for stimulating discussions. This study was supported in part by
the Polish Committee for Scientific Research, grants KBN-2P03B 120 23 and PBZ-KBN-044/P03-2001 (M.K.).

\Figures

\end{document}